\documentclass[12pt]{article}
\usepackage{}
\usepackage{graphicx}
\usepackage{amssymb}
\usepackage{amsmath}
\usepackage{bm}
\usepackage{indentfirst}
\usepackage{cite}

\allowdisplaybreaks
\begin{document}
\title{Dissociation of Large-$p_{\rm{T}}$ Prompt $J/\psi$ 
\\ Produced in Pb-Pb Collisions at the LHC}
\author{Shi-Tao Ji$^{1}$, Xiao-Ming Xu$^{1}$ and H. J. Weber$^{2}$}
\date{}
\maketitle \vspace{-1cm}
\centerline{$^{1}$Department of Physics, Shanghai University, Baoshan,
Shanghai 200444, China}
\centerline{$^{2}$Department of Physics, University of Virginia, 
Charlottesville, VA 22904, USA}

\begin{abstract}
A collision of a light meson and a charmonium produces quarks and
antiquarks first, then the charm quark fragments into a charmed
hadron, and finally three or more mesons are produced. This is the
mechanism that we consider and propose to understand the
prompt-$J/\psi$ nuclear modification factor in the range of
$J/\psi$ transverse momentum from 10 GeV/$c$ to 18 
GeV/$c$, as measured by the CMS Collaboration and the ATLAS 
Collaboration. Unpolarized cross sections are
derived and calculated for the reactions $\pi + {\rm charmonium}
\to H_c +X$ with $H_c$ being $D^+$, $D^0$, $D^+_s$, $D^{*+}$, or
$D^{*0}$. Numerical cross sections are parametrized and used to
calculate the dissociation rate of charmonium in the interaction
with pions in hadronic matter. The momentum dependence of the rate
obtained is so as to lead to a decreasing charmonium nuclear
modification factor with increasing transverse momentum.
\end{abstract}



\vspace{0.5cm}
\leftline{\bf 1. Introduction}
\vspace{0.5cm}

Even though no hadronic meson beam collision experiments are possible,
theoretical efforts to understand hadron-charmonium collisions have
been made.
With the idea that a gluon with an energy larger than the binding
energy of a heavy quarkonium breaks the quarkonium, the short-distance
approach \cite{Peskin,KS,AGGA}
gives very small hadron-quarkonium cross sections near the
threshold energy and increasing cross section at large center-of-mass
energy $\sqrt s$ of the hadron and the quarkonium. 
This approach does
not specify what are the final hadrons formed in the breakup of the
heavy quarkonium. 

Through exchange of charmed mesons a light hadron breaks
$J/\psi$. In the meson-exchange approach \cite{MM,HG,LK,OSL,NNR,MPPR,BG}
the cross section for any endothermic reaction increases rapidly as
$\sqrt s$ increases from the threshold energy. Haglin and Gale showed
that the $\pi + J/\psi$ ($\rho + J/\psi$) total inelastic cross
section would reach an extremely large value of roughly 100 mb (200
mb) at $\sqrt{s} =6$ GeV. 
Similar large cross sections have been reported by Oh, Song and
Lee. The incorrect magnitude of cross sections comes from pointlike
hadron vertices between which the exchange of heavy mesons leads to an
interaction too short-ranged to allow the interacting particles to 
initiate the required $J/\psi$ dissociation reactions. To get reasonable cross 
sections and to justify the meson-exchange approach, hadronic form factors
are inserted in three-meson and four-meson vertices.
When the center-of-mass energy of the hadron and $J/\psi$ rises far
away from threshold, the increase or decrease of the cross section
depends on the choice of the form factors. At present, the 
meson-exchange
approach deals with only 2-to-2 hadron-$J/\psi$ reactions.

We may use quark potential models in the Born approximation to
study hadron-charmonium dissociation which is governed by
interchange of the quark in the hadron and the charm quark in the
charmonium \cite{MBQ,WSB,BSWX,ZX,JSX}. It is concluded in the 
quark-interchange approach that the dissociation
cross section increases rapidly from 0 for an
endothermic reaction or decreases rapidly from infinity for an 
exothermic reaction while
$\sqrt s$ increases from the threshold energy. No matter which 
reaction is relevant, endothermic or exothermic, the cross
section is very small when $\sqrt s$ is larger than the threshold
energy plus 1 GeV. Now we raise the question, do the quark potential 
models really give the very small cross sections at large 
$\sqrt s$? The answer is no, but we
mention that the quark-interchange mechanism leads to the 2-to-2 
hadron-charmonium reactions with the above results. Therefore, we
need to think about a
new mechanism that provides appreciable hadron-charmonium
dissociation cross sections at large $\sqrt s$. This is the subject
of the present work.

Charmonium dissociation causes the nuclear modification of
charmonia. The change of the nuclear modification of $J/\psi$ as a
function of its transverse momentum $p_T$ is very interesting in
relativistic heavy-ion collisions. For nucleus-nucleus collisions
at $\sqrt{s_{NN}}=200$ GeV the prompt-$J/\psi$ transverse momentum
measured by the PHENIX Collaboration \cite{PHENIX} and the STAR
Collaboration \cite{STAR} is well below 10 GeV/$c$.  
For Pb-Pb collisions at
the Large Hadron Collider (LHC) the prompt-$J/\psi$ transverse
momentum measured by the CMS Collaboration at 
$\sqrt {s_{NN}}=2.76$ TeV \cite{CMS} and the ATLAS 
Collaboration at $\sqrt {s_{NN}}=5.02$ TeV \cite{ATLAS} can be
beyond 10 GeV/$c$. Both data show a suppression pattern of the prompt
$J/\psi$. The $p_T$ dependence of the ATLAS data is that the nuclear
modification factor stays unchanged from 10 GeV/$c$ to 18 GeV/$c$ and
increases slowly from 18 GeV/$c$ to 40 GeV/$c$. The contribution
of the nuclear absorption to the nuclear modification factor
is expected to be negligible due to the
very small $c\bar c$ size produced in a parton-parton collision
and the very short crossing time of the $c\bar c$ pair through
colliding nuclei at the LHC.
The meson-charmonium dissociation induced by
quark interchange in hadronic matter is completely negligible in
the suppression of $J/\psi$ with a momentum beyond 10 GeV/$c$
\cite{LJX}. Recombination of charm quarks and charm antiquarks in
a quark-gluon plasma does not contribute to the nuclear
modification factor beyond 10 GeV/$c$ \cite{Xu1999,DR}. The $J/\psi$
nuclear modification at such large transverse momenta are 
caused by other factors, for example, nuclear parton shadowing in 
the colliding nuclei \cite{Arneodo,EKS,LW}, $J/\psi$ 
dissociation in collisions with gluons in the quark-gluon plasma 
\cite{KS,XKSW,ZXXZ}, decreasing dissociation temperature with
increasing $J/\psi$ velocity \cite{LRW}, and radiative energy
loss of a gluon traveling through the plasma before gluon
fragmentation into a $J/\psi$ meson. Yet, dilating formation
time of $J/\psi$ with increasing momentum observed in the plasma 
rest frame reduces the $J/\psi$ suppression
\cite{Xu1999}. We denote by $R_{AA}^{\prime}$ the nuclear 
modification factor resulting from the above factors together with
the formation-time effect. 
For the $J/\psi$ meson with a momentum larger than
10 GeV/$c$, in the present work, we consider that three or more 
mesons are produced in the collision of a light meson and a
charmonium. One of the produced mesons is a charmed meson.
We estimate cross sections for $\pi + J/\psi \to H_c +X$ with $H_c$
being $D^+$, $D^0$, $D^+_s$, $D^{*+}$, or $D^{*0}$ and $X$
representing two or more mesons, and calculate the
dissociation rate of $J/\psi$ in the interaction with $\pi$ in
hadronic matter. The $J/\psi$ meson with the momentum 20 GeV/$c$ has
the dissociation rate 0.0108 $c$/fm at a temperature of 0.95 times
the critical temperature $T_{\rm c}$. If the lifetime of hadronic 
matter in longitudinal expansion with a freeze-out temperature of
$0.7T_{\rm c}$ is 5-10 fm/$c$, inclusion
of the $\pi + J/\psi$ reaction leads to the nuclear modification
factor which is 0.973-0.946 times $R_{AA}^{\prime}$. Therefore, 
the meson-charmonium dissociation mechanism
proposed here is a source for the $J/\psi$ suppression at momenta
beyond 10 GeV/$c$. This is in contrast to the quark-interchange 
mechanism which produces two charmed mesons in meson-charmonium
dissociation and works well for low-momentum charmonia 
\cite{MBQ,WSB,BSWX,ZX,JSX}.
  
In Section 2 the mechanism is given and relevant formulas are
presented. In Section 3 we show $\pi$-charmonium dissociation
cross sections generated by the mechanism and the dissociation rate of
the charmonium in the interaction with the pion in hadronic matter.
Relevant discussions are given. Section 4 contains the summary.

\vspace{0.5cm}
\leftline{\bf 2. Formulas}
\vspace{0.5cm}

We consider the reaction $A(q\bar{q})+B(c\bar{c})\to
q+\bar{q}+c+\bar{c}\to H_c+X$, which means that a collision of
meson $A$ and meson $B$ produces quarks and antiquarks ($q$,
$\bar{q}$, $c$, and $\bar{c}$), the charm quark fragments into
hadron $H_c$, and $q$, $\bar{q}$, and $\bar{c}$ give rise to two
or more mesons. The symbol $X$ actually represents two or more
mesons of which one meson contains $\bar{c}$. The reaction breaks
the charmonium, i.e., meson $B$. The differential cross section
for the reaction is
\begin{eqnarray}
d\sigma &=& \frac{(2\pi)^4|{\cal M}_{\rm fi}|^2}
{4\sqrt{(P_{A}\cdot P_{B})^2-m_{A}^2m_{B}^2}}
\frac{d^3p^{\prime}_{q}}{(2\pi)^32E^{\prime}_{q}}
\frac{d^3p^{\prime}_{\bar{q}}}{(2\pi)^32E^{\prime}_{\bar{q}}}
\frac{d^3p^{\prime}_{c}}{(2\pi)^32E^{\prime}_{c}}
\frac{d^3p^{\prime}_{\bar{c}}}{(2\pi)^32E^{\prime}_{\bar{c}}}
     \nonumber\\
&&\times dzD_{c}^{H_{c}}(z,\mu^2)
\delta^4(P_{A}+P_{B}-p^{\prime}_{q}-p^{\prime}_{\bar{q}}-p^{\prime}_{c}-p^{\prime}_{\bar{c}}),
\end{eqnarray}
where $p^{\prime}_i=(E^{\prime}_i,\vec{p}^{~\prime}_i)$
$(i=q,\bar{q},c,\bar{c})$ are the four-momenta of $q$, $\bar{q}$,
$c$, and $\bar{c}$; $m_{A}$ ($m_{B}$) and $P_{A}$ ($P_{B}$) are
the mass and four-momentum of meson $A$ ($B$), respectively;
$D_{c}^{H_{c}}(z,\mu^2)$ is the $c \to H_c$ fragmentation function
at the factorization scale $\mu$; $z$ is the fraction of energy
passed on from quark $c$ to hadron $H_c$; ${\cal M}_{\rm fi}$ is
the transition amplitude for $A+B\to q+\bar{q}+c+\bar{c}$.

With the Mandelstam variable $s=(P_A+P_B)^2$, integration over
$\vec{p}^{~\prime}_{\bar{q}}$ and $|\vec{p}^{~\prime}_q|$ yields
\begin{eqnarray}
d\sigma &=& \frac{1}{32(2\pi)^8\sqrt{[s-(m_{A}+m_{B})^2][s-(m_{A}-m_{B})^2]}}
d\Omega^{\prime}_{q}\frac{d^3p^{\prime}_{c}}{E^{\prime}_{c}}
\frac{d^3p^{\prime}_{\bar{c}}}{E^{\prime}_{\bar{c}}}
     \nonumber\\
&&\times dzD_{c}^{H_{c}}(z,\mu^2)
\frac{|\vec{p}^{~\prime}_{q}|_0^2| {\cal M}_{\rm
fi}|^2}{\left||\vec{p}^{~\prime}_{q}|_0E^{\prime}_{\bar{q}}+
(|\vec{p}^{~\prime}_{q}|_0-|\vec{P}_{A}+\vec{P}_{B}-\vec{p}^{~\prime}_{c}-\vec{p}^{~\prime}_{\bar{c}}|\cos\Theta)
E^{\prime}_{q}\right|},
\end{eqnarray}
where $\Theta$ is the angle between $\vec{p}^{~\prime}_{q}$ and
$\vec{P}_{A}+\vec{P}_{B}-\vec{p}^{~\prime}_{c}-\vec{p}^{~\prime}_{\bar{c}}$,
and $d\Omega^{\prime}_{q}$ is the solid angle centered about the
direction of $\vec{p}^{~\prime}_{q}$. $|\vec{p}^{~\prime}_{q}|_0$
is determined by the energy conservation,
\begin{eqnarray}
E_{A}+E_{B}-E^{\prime}_{q}-E^{\prime}_{\bar{q}}-E^{\prime}_{c}-E^{\prime}_{\bar{c}}=0,
\end{eqnarray}
where $E_A$ and $E_B$ are the energies of mesons $A$ and $B$,
respectively.

We calculate the cross section in the center-of-momentum frame of
mesons $A$ and $B$. Denote the maxima of $|\vec{p}^{~\prime}_c|$
and $|\vec{p}^{~\prime}_{\bar{c}}|$ by $p^{\prime}_{c\rm{max}}$
and $p^{\prime}_{\bar{c}\rm{max}}$, respectively. We obtain
$p^{\prime}_{c\rm{max}}$ and $p^{\prime}_{\bar{c}\rm{max}}$ by
setting $\vec{p}^{~\prime}_q=\vec{p}^{~\prime}_{\bar{q}}=0$, and
they satisfy
\begin{eqnarray}
\sqrt{s}=\sqrt{m_c^2+p^{\prime2}_{c\rm{max}}}+\sqrt{m_{\bar{c}}^2+p^{\prime2}_{\bar{c}\rm{max}}}+
m_q+m_{\bar{q}},
\end{eqnarray}
where $m_i$ $(i=q,\bar{q},c,\bar{c})$ are the masses of $q$,
$\bar{q}$, $c$, and $\bar{c}$. The maxima are
\begin{eqnarray}
p^{\prime}_{\bar{c}\rm{max}}&=&p^{\prime}_{c\rm{max}}=\frac{1}{2(\sqrt{s}-m_q-m_{\bar{q}})}
[(\sqrt{s}-m_q-m_{\bar{q}}-m_c-m_{\bar{c}})
     \nonumber\\
&&(\sqrt{s}-m_q-m_{\bar{q}}+m_c-m_{\bar{c}})(\sqrt{s}-m_q-m_{\bar{q}}-m_c+m_{\bar{c}})
\nonumber\\
&&(\sqrt{s}-m_q-m_{\bar{q}}+m_c+m_{\bar{c}})]^\frac{1}{2}.
\end{eqnarray}
Denote the minimum of $|\vec{p}^{~\prime}_c|$ by $p^{\prime}_{c\rm{min}}$. The
fragmentation of the charm quark into hadron $H_c$ requires
$\sqrt{m^2_c+p^{\prime2}_{c\rm{min}}}=m_{H_c}$, which gives
\begin{eqnarray}
p^{\prime}_{c\rm{min}}=\sqrt{{m^2_{H_c}}-m^2_c},
\end{eqnarray}
where $m_{H_c}$ is the mass of hadron $H_c$.

The cross section for $A(q\bar{q})+B(c\bar{c}) \to q+\bar{q}+c+\bar{c} \to H_{c}+X$ via $c \to H_{c}$ is
\begin{eqnarray}
\sigma(\sqrt {s}, T) &=&
\frac{1}{32(2\pi)^8\sqrt{[s-(m_{A}+m_{B})^2][s-(m_{A}-m_{B})^2]}}
\int d\Omega^{\prime}_{q}\int^{p^{\prime}_{c\rm
{max}}}_{p^{\prime}_{c\rm{min}}}
\frac{d^3p^{\prime}_{c}}{E^{\prime}_{c}}
     \nonumber\\
&&
\times\int^{p^{\prime}_{\bar{c}\rm{max}}}_0
\frac{d^3p^{\prime}_{\bar{c}}}{E^{\prime}_{\bar{c}}}
\int^1_{m_{H_{c}}/E^{\prime}_{c}}dzD_{c}^{H_{c}}(z,\mu^2)
     \nonumber\\
&&
\times\frac{|\vec{p}^{~\prime}_{q}|_0^2|{\cal M}_{\rm fi}|^2}{\left||\vec{p}^{~\prime}_{q}|_0E'_{\bar{q}}+(|\vec{p}^{~\prime}_{q}|_0-
|\vec{P}_{A}+\vec{P}_{B}-\vec{p}^{~\prime}_{c}-\vec{p}^{~\prime}_{\bar{c}}|\cos\Theta)E^{\prime}_{q}\right|}.
\label{eq:cs}
\end{eqnarray}
In Eq. (\ref{eq:cs}) the transition amplitude is given by
\begin{eqnarray}
{\cal M}_{\rm fi} &=& \sqrt{2E_{A}2E_{B}2E'_{q}2E'_{\bar{q}}2E'_{c}2E'_{\bar{c}}}
     \nonumber\\
&&\times \left[ V_{q\bar{c}}(\vec{Q})
\psi_{A}(\vec{p}^{~\prime}_{q\bar{q}}-\frac{m_{\bar{q}}}{m_q+m_{\bar{q}}}\vec{Q})
\psi_{B}(\vec{p}^{~\prime}_{c\bar{c}}-\frac{m_c}{m_c+m_{\bar{c}}}\vec{Q})
\right.
     \nonumber\\
&&+V_{\bar{q}c}(\vec{Q})
\psi_{A}(\vec{p}^{~\prime}_{q\bar{q}}+\frac{m_q}{m_q+m_{\bar{q}}}\vec{Q})
\psi_{B}(\vec{p}^{~\prime}_{c\bar{c}}+\frac{m_{\bar{c}}}{m_c+m_{\bar{c}}}\vec{Q})
     \nonumber\\
&&+V_{qc}(\vec{Q})
\psi_{A}(\vec{p}^{~\prime}_{q\bar{q}}-\frac{m_{\bar{q}}}{m_q+m_{\bar{q}}}\vec{Q})
\psi_{B}(\vec{p}^{~\prime}_{c\bar{c}}+\frac{m_{\bar{c}}}{m_c+m_{\bar{c}}}\vec{Q})
     \nonumber\\
&& \left. +V_{\bar{q}\bar{c}}(\vec{Q})
\psi_{A}(\vec{p}^{~\prime}_{q\bar{q}}+\frac{m_q}{m_q+m_{\bar{q}}}\vec{Q})
\psi_{B}(\vec{p}^{~\prime}_{c\bar{c}}-\frac{m_c}{m_c+m_{\bar{c}}}\vec{Q})
\right] ,
\end{eqnarray}
where $\vec{Q}$ is the gluon three-dimensional momentum;
$\vec{p}^{~\prime}_{ij}$ ($i,j=q,\bar{q},c,\bar{c}$) is the
relative quark momentum; $\psi_{A}$ ($\psi_{B}$) represents the
product of color, spin, flavor, and relative-motion wave functions
of the quark and antiquark inside meson $A$ ($B$), and satisfies
$\int\frac{d^3p_{q\bar{q}}}{(2\pi)^3}\psi^+_A(\vec{p}_{q\bar{q}})\psi_A(\vec{p}_{q\bar{q}})=
\int\frac{d^3p_{c\bar{c}}}{(2\pi)^3}\psi^+_B(\vec{p}_{c\bar{c}})\psi_B(\vec{p}_{c\bar{c}})=1$,
where $\vec{p}_{q\bar q}$ ($\vec{p}_{c\bar c}$) is the relative
momentum of the quark and antiquark inside meson $A$ ($B$).

The fragmentation functions are solutions of the 
Dokshitzer-Gribov-Lipatov-Altarelli-Paris (DGLAP) evolution
equations \cite{KKKS}. The starting point $\mu_0$ for the DGLAP 
evolution in
the factorization scale $\mu$ is taken to be the charm quark
mass, i.e., $\mu_0=m_c$. The charm-quark fragmentation functions at
$\mu_0$ are given in Ref. \cite{KK} at
next-to-leading order in the modified minimal-subtraction
factorization scheme by fitting the $e^+ e^-$ data taken by the
OPAL Collaboration at the CERN Large Electron-Positron Collider:
\begin{equation}
D_c^{D^+}(z,\mu_0^2)=0.266 \frac {z(1-z)^2}{[(1-z)^2+0.108z]^2},
\end{equation}
\begin{equation}
D_c^{D^0}(z,\mu_0^2)=0.781 \frac {z(1-z)^2}{[(1-z)^2+0.119z]^2},
\end{equation}
\begin{equation}
D_c^{D_s^+}(z,\mu_0^2)=0.0381\frac {z(1-z)^2}{[(1-z)^2+0.0269z]^2},
\end{equation}
\begin{equation}
D_c^{D^{*+}}(z,\mu_0^2)=0.192\frac {z(1-z)^2}{[(1-z)^2+0.0665z]^2}.
\end{equation}
As in Ref. \cite{KKKS} we set $\mu=\sqrt s$. The fragmentation
functions are assumed to be universal and are thus used in 
hadron-charmonium reactions.

Denote the orbital angular momentum and the spin of meson $A$ ($B$) by $L_A$ 
($L_B$) and $S_A$ ($S_B$),
respectively. Let $L_{Bz}$ be the magnetic projection quantum number of $L_B$. 
The unpolarized cross section is
\begin{eqnarray}
\sigma^{\rm{unpol}}(\sqrt{s},T)=\frac{1}{(2S_A+1)(2S_B+1)(2L_B+1)}\sum
\limits_{L_{Bz}SS_{q+\bar{q}}S_{c+\bar{c}}}
(2S+1)\sigma(\sqrt{s},T),
\end{eqnarray}
which holds true for the three cases: $L_A=0$, $L_B=0$; $L_A=0$, $L_B\neq0$, 
$S_A=0$; $L_A=0$, $L_B=1$, $S_A=1$,
$S_B=1$. The total spin $S$ of mesons $A$ and $B$ satisfies 
$|S_A-S_B|\leq{S}\leq S_A+S_B$ and
$|S_{q+\bar{q}}-S_{c+\bar{c}}|\leq{S}\leq|S_{q+\bar{q}}+S_{c+\bar{c}}|$, 
where $S_{q+\bar{q}}$
($S_{c+\bar{c}}$) is the total spin of $q$ and $\bar{q}$ ($c$ and $\bar{c}$).

The potential $V_{ab}$ used in the transition amplitude is the
Fourier transform of the following potential in coordinate space
\cite{ZX},
\begin{eqnarray}
V_{ab}(\vec {r}) = V_{\rm{si}}(\vec {r})+V_{\rm{ss}}(\vec {r}),
\end{eqnarray}
where $\vec {r}$ is the relative coordinate of constituents $a$
and $b$, $V_{\rm{si}}$ the central spin-independent potential, and
$V_{\rm{ss}}$ the spin-spin interaction. The spin-independent
potential depends on temperature $T$ and below the critical
temperature $T_{\rm c}=0.175$ GeV is given by
\begin{equation}
V_{\rm {si}}(\vec {r})=
-\frac {\vec {\lambda}_a}{2} \cdot \frac {\vec{\lambda}_b}{2}
\frac{3}{4} D \left[ 1.3- \left( \frac {T}{T_{\rm c}} \right)^4 \right]
\tanh (Ar)
+ \frac {\vec {\lambda}_a}{2} \cdot \frac {\vec {\lambda}_b}{2}
\frac {6\pi}{25} \frac {v(\lambda r)}{r} \exp (-Er),
\end{equation}
where $D=0. 7$ GeV, $A=1.5[0.75+0.25 (T/{T_{\rm c}})^{10}]^6$ GeV,
$E=0. 6$ GeV, and $\lambda=\sqrt{3b_0/16\pi^2 \alpha'}$ in which
$\alpha'=1.04$ GeV$^{-2}$ and $b_{0}=11-\frac{2}{3}N_{f}$ with the
quark flavor number $N_{f}=4$. $\vec {\lambda}_a$ are the
Gell-Mann matrices for the color generators of constituent quark
or antiquark labeled as $a$. The dimensionless function $v(x)$ is
given by Buchm\"{u}ller and Tye \cite{BT}. The short-distance part
of the spin-independent potential originates from one-gluon
exchange plus perturbative one- and two-loop corrections. The
intermediate-distance and large-distance part of the
spin-independent potential fits well the numerical potential which
was obtained in the lattice gauge calculations \cite{KLP} and 
was assumed to be the free energy. At large
distances the spin-independent potential is independent of the
relative coordinate and obviously exhibits a plateau at $T/T_{\rm
c} > 0.55$. The plateau height decreases with increasing
temperature. This means that confinement becomes weaker and
weaker.

In this paragraph we mention six references which show that the
quark-antiquark free energy can not be taken as the 
quark-antiquark potential when the temperature is above the
critical temperature. The quark-antiquark free energy 
$F_{Q\bar Q}$ is defined as the 
quark-antiquark internal energy $U_{Q\bar Q}$ minus the product
of the temperature and the quark-antiquark entropy $S_{Q\bar Q}$. 
The internal energy of a quark and an antiquark at rest is the
quark-antiquark potential. The free energy of a heavy
quark-antiquark pair is related to the Polyakov loop
correlation functions and can be obtained from lattice gauge
calculations. It is real! Since the lattice calculations provide
the internal energy $U(T,r)$ that includes the internal energy of
the heavy quark-antiquark pair and the gluon internal energy
difference $U_g(T,r)-U_{g0}(T)$ where $U_g(T,r)$ and $U_{g0}(T)$
correspond to the gluon internal energies in the presence and the
absence of the heavy quark-antiquark pair, respectively, it was
proposed by Wong \cite{Wong2005} that the heavy quark-antiquark
potential should be $U(T,r)-[U_g(T,r)-U_{g0}(T)]$ and may be
taken as a linear combination of $F_{Q\bar Q}$ and $U(T,r)$.
Authors in Ref. \cite{LPTR} have derived a static quark-antiquark
potential in a deconfined medium by defining a suitable
gauge-invariant Green's function and computing it to first
non-trivial order in Hard Thermal Loop resummed perturbation
theory. The potential consists of a real part and an imaginary
part. Authors in Ref. \cite{BGVP} have also derived a static
quark-antiquark potental in the deconfined medium in potential 
nonrelativistic QCD. The potential has various expressions that
depend on the relation among
the temperature, the Debye mass, and the inverse of
the quark-antiquark distance. Authors in Refs. 
\cite{BKR14,BKR16,BBP}
have obtained a quark-antiquark potential from the spectral
function of the thermal Wilson loop in Coulomb gauge
in dynamical lattice QCD with $u$, $d$, and $s$ flavors. The 
imaginary part of the potential is comparable with the result of
the Hard Thermal Loop resummed perturbation theory.

In this paragraph we mention two references which show that the
quark-antiquark free energy can be taken as the quark-antiquark
potential when the temperature is above the critical temperature.
Using the free energy from lattice calculations
as the potential of a charm quark and a charm antiquark in the 
Schr\"odinger equation correctly describes the nonrelativistic 
$J/\psi$ wave function and reproduces the $J/\psi$ mass 
from the QCD sum rule in the vicinity of the critical temperature
\cite{LMSK}. 
The entropic force is the derivative of the entropy with respect
to the quark-antiquark distance when the temperature is fixed.
The stronger increase of the internal energy with increasing
quark-antiquark distance is compensated by an equally strong
contribution from the repulsive entropic force. Thus, it is  
suggested by Satz that the potential is the
free energy \cite{Satz}.

In this paragraph we indicate that the
quark-antiquark free energy can be taken as the quark-antiquark
potential when the temperature is below the critical temperature.
Let the quark-antiquark potential obtained from the spectral 
function of the thermal Wilson loop be compared to the 
quark-antiquark free energy. It has been realized that the real
part of the potential is close to the free energy at $T<T_{\rm c}$
and the imaginary part is negligible \cite{BKR14,BKR16,RHS,BR}. 
In hadronic matter $TS_{Q\bar Q}$ is quite small in comparison with
the free energy, and the quark-antiquark internal energy, 
i.e., the potential, then approximately equals
the free energy \cite{SX}.

Denote by $\vec {s}_a$ the spin of constituent $a$. The spin-spin
interaction with relativistic effects \cite{BS,GI} is
\cite{ZX,Xu2002}
\begin{eqnarray}
V_{\rm ss}(\vec {r})=
- \frac {\vec {\lambda}_a}{2} \cdot \frac {\vec {\lambda}_b}{2}
\frac {16\pi^2}{25}\frac{d^3}{\pi^{3/2}}\exp(-d^2r^2) \frac {\vec {s}_a \cdot \vec
{s} _b} {m_am_b}
+ \frac {\vec {\lambda}_a}{2} \cdot \frac {\vec {\lambda}_b}{2}
  \frac {4\pi}{25} \frac {1} {r}
\frac {d^2v(\lambda r)}{dr^2} \frac {\vec {s}_a \cdot \vec {s}_b}{m_am_b},\label{eq:vss}
\end{eqnarray}
of which the flavor dependence is relevant to quark masses as
shown in $1/m_am_b$ and $d$,
\begin{eqnarray}
d^2=\sigma_{0}^2\left[\frac{1}{2}+\frac{1}{2}\left(\frac{4m_a m_b}{(m_a+m_b)^2}\right)^4\right]
+\sigma_{1}^2\left(\frac{2m_am_b}{m_a+m_b}\right)^2,
\label{eq:d}
\end{eqnarray}
where $\sigma_0=0.15$ GeV and $\sigma_1=0.705$.

Solving the Schr\"{o}dinger equation with the central
spin-independent potential plus the spin-spin interaction, we
obtain meson masses and quark-antiquark relative-motion wave
functions with the charm-quark mass, the up-quark mass, and the
strange-quark mass being 1.51 GeV, 0.32 GeV, and 0.5 GeV,
respectively. The experimental masses of $\pi$, $\rho$, $K$,
$K^*$, $\eta$, $\phi$, $J/\psi$, $\psi'$, $\chi_c$, $D$, $D^*$,
$D_s$, and $D^*_s$ mesons \cite{PDG} and the experimental data of
$S$-wave $I=$ 2 elastic phase shifts for $\pi\pi$ scattering in
vacuum \cite{Colton} are reproduced with $V_{ab}(\vec {r})$ at
$T=$ 0 and the quark-antiquark relative-motion wave functions.

In the mechanism shown above we assume that the interactions among
$q$, $\bar q$, $c$, and $\bar c$ are neglected, and the 
fragmentation of the charm quark into hadrons is related to
quark-antiquark pairs created from the color field around the
charm quark, which is similar to the scenario
of Feynman and Field \cite{FF}. We do not address how the
mesons represented by $X$ are formed, and also
neglect any temperature dependence of the fragmentation.

\vspace{0.5cm}
\leftline{\bf 3. Numerical results and discussions}
\vspace{0.5cm}

We consider the charmonium dissociation $\pi+J/\psi\to H_c+X$,
$\pi+\psi'\to H_c+X$, and $\pi+\chi_c\to H_c+X$ with $H_c=D^+$,
$D^0$, $D_s^+$, $D^{*+}$, or $D^{*0}$. This amounts to 15 reaction
channels. We solve the Schr\"{o}dinger equation with the potential
in Eq. (14) to get temperature-dependent quark-antiquark
relative-motion wave functions of $\pi$, $J/\psi$, $\psi'$, and
$\chi_c$, calculate the transition amplitude with the wave
functions and the potential, and obtain unpolarized cross sections
from the transition amplitude and the charm-quark fragmentation
functions provided in Ref. \cite{KK}. The unpolarized cross
sections for  $\pi+J/\psi\to H_c+X$ at the temperatures
$T/T_{\rm{c}}=$ 0, 0.65, 0.75, 0.85, 0.9, and 0.95 are shown in
Figs. 1-4. Because the $D^{*0}$ fragmentation function is unknown,
we assume that it is the same as the $D^{*+}$ fragmentation
function. The cross section for $\pi+J/\psi\to D^{*0}+X$ then
equals the cross section for $\pi+J/\psi\to D^{*+}+X$ and is not
shown.

All the reactions are endothermic as shown in Figs. 1-4. When the
difference between $\sqrt{s}$ and the threshold energy is less
than 1 GeV, the unpolarized cross section is negligible. With
increasing $\sqrt{s}$ the cross section increases slowly. At a given 
temperature and difference between $\sqrt{s}$ and the threshold
energy, the cross section for the production of $D^0$ is largest,
the cross section for the production of $D_s^+$ is smallest, and
the cross section for $\pi+\psi'\to H_c+X$ or $\pi+\chi_c\to
H_c+X$ is larger than the one for $\pi+J/\psi\to H_c+X$. Since the
cross-section curves for $\pi+\psi'\to H_c+X$ and $\pi+\chi_c\to
H_c+X$ look similar to the ones for $\pi+J/\psi\to H_c+X$, the
former are not shown here.

In Eq. (7) the integration over $\mid \vec{p}_c^{~\prime} \mid$
($\mid \vec{p}_{\bar c}^{~\prime} \mid$) has the lower limit
$p^\prime_{c\rm min}$ (0) and the upper limit 
$p^\prime_{c\rm max}$ ($p^\prime_{\bar{c}\rm max}$).
$p^\prime_{c\rm min}$ given in Eq. (6) is independent of 
$\sqrt s$. While $\sqrt s$ increases, $p^\prime_{c\rm max}$ and
$p^\prime_{\bar{c}\rm max}$ given in Eq. (5)
increase. Hence, the phase space of
the produced charm quark and the produced charm antiquark in
$A+B \to q+\bar{q}+c+\bar{c}$ becomes
larger and larger. While $\sqrt s$ increases, $\sqrt {E_AE_B}$ in
Eq. (8) also increases, but the factor 
$1/\sqrt{[s-(m_{A}+m_{B})^2][s-(m_{A}-m_{B})^2]}$ in Eq. (7)
decreases. Therefore, while $\sqrt s$ increases from the threshold
energy, the cross section increases first, then reaches a maximum,
and eventually decreases slowly.

In vacuum the central spin-independent potential is given by the
Buchm\"uller-Tye potential,
\begin{equation}
V_{\rm {BT}}(\vec {r})=
-\frac {\vec {\lambda}_a}{2} \cdot \frac {\vec{\lambda}_b}{2}
\frac{3}{4} kr
+ \frac {\vec {\lambda}_a}{2} \cdot \frac {\vec {\lambda}_b}{2}
\frac {6\pi}{25} \frac {v(\lambda r)}{r} ,
\end{equation}
with $k=1/(2\pi\alpha')$. The quark potential in vacuum is
\begin{equation}
V_{\rm {vac}}(\vec {r})=V_{\rm {BT}}(\vec {r})+V_{\rm {ss}}(\vec {r}).
\end{equation}
Using the potential $V_{\rm {vac}}(\vec {r})$ in the transition 
amplitude, we get numerical unpolarized cross sections by  Eq. (13).
Dotted curves
in Fig. 8 show the cross sections for $\pi+J/\psi\to H_c+X$ 
which are compared with the solid
curves that stand for the unpolarized cross sections corresponding
to the potential in Eq. (14) at $T=0$. The four solid curves in Fig.
8 are actually the four solid curves in Figs. 1-4.
The dotted curves are quite close to the solid curves. 
The situation is similar with respect to the
cross sections for $\pi+\psi'\to H_c+X$ and $\pi+\chi_c \to H_c+X$,
which are not shown. Thus, 
the cross sections corresponding to the quark potential in vacuum
are quite close to the ones corresponding to the potential in
Eq. (14) at $T=0$. Combining Fig. 8 with Figs. 1-4, every 
cross-section curve at a nonzero temperature in Figs. 1-4
should intersect a
dotted curve in Fig. 8 at a value of $\sqrt s$. Below this value
the cross section related to Eq. (14) is larger than the cross
section related to Eq. (19) while above this value the
former is smaller than the latter. At $\sqrt {s}=7$ GeV  (11 GeV)
the cross section for $\pi+J/\psi\to D^++X$, $\pi+J/\psi\to D^0+X$,
$\pi+J/\psi\to D_s^++X$, and $\pi+J/\psi\to D^{*+}+X$ at 
$T/T_c=0.95$ is 1.97, 1.94, 1.54, and 3.17 (0.77, 0.79, 0.68, and 
0.75) times the cross section corresponding to
the quark potential in vacuum, respectively.  The
medium effect on the reactions is obvious at $T/T_c=0.95$.

A cross section of about 1 mb for $\pi + J/\psi$ dissociation at
$\sqrt{s}=10$ GeV was obtained in the short-distance approach 
\cite{AGGA}. The usual monopole form with a cutoff parameter is
taken as the form factor inserted in three-meson
vertices, and the product of two monopole forms is taken as the
form factor inserted in four-meson vertices in the 
meson-exchange approach. Starting from an effective meson Lagrangian,
a cross section of about 3.6 mb for $\pi + J/\psi$ dissociation at
$\sqrt{s}=5$ GeV is given in Ref. \cite{LK} with the cutoff
parameter 1 GeV. Adding anomalous parity terms to the Lagrangian, a
cross section of about 6.4 mb at $\sqrt{s}=5$ GeV is given in Ref.
\cite{OSL}. While mesonic form factors 
were calculated from the underlying quark structure, $J/\psi$
dissociation in collisions with $\pi$ and $\rho$ mesons were studied
in Ref. \cite{BG} in an extended Nambu-Jona-Lasinio model. The
$\pi + J/\psi$ dissociation cross section has a peak value of about
0.83 mb near the threshold energy of $\pi + J/\psi$ dissociation. In
the quark-interchange mechanism and in the
quark model that possesses the color Coulomb, spin-spin hyperfine,
and linear confining interactions, the cross section obtained in Ref.
\cite{BSWX} has a peak value of 1.41 mb near threshold. The five
values (1 mb, 3.6 mb, 6.4 mb, 0.83 mb, and 1.41 mb) characterize the 
$\pi + J/\psi$ dissociation cross sections obtained in the 
short-distance approach, in the meson-exchange approach, and in the
quark-interchange approach. In the present work the cross section 
for $\pi J/\psi \to D^+X + D^0X + D_s^+X + D^{*+}X + D^{*0}X$ is, 
for example, 1.46 mb at $\sqrt{s} =10$ GeV corresponding to the
potential in Eq. (14) at $T=0$, or 1.49 mb corresponding to the
potential in Eq. (19). The two values, 1.46 mb and 1.49 mb, are
comparable to the above five values. Therefore, the cross section
for $\pi + J/\psi$ dissociation due to the present mechanism is
not very small at large $\sqrt{s}$, 
but depends on the selection among quark potential models.

For the convenient use of the unpolarized cross sections shown
in Figs. 1-4, they are parametrized as
\begin{eqnarray}
\sigma^{\rm unpol}(\sqrt {s},T)
&=&a_1 \left( \frac {\sqrt {s} -\sqrt {s_0}} {b_1} \right)^{c_1}
\exp \left[ c_1 \left( 1-\frac {\sqrt {s} -\sqrt {s_0}} {b_1} \right) \right]
\nonumber \\
&&+ a_2 \left( \frac {\sqrt {s} -\sqrt {s_0}} {b_2} \right)^{c_2}
\exp \left[ c_2 \left( 1-\frac {\sqrt {s} -\sqrt {s_0}} {b_2} \right) \right],
\end{eqnarray}
where $\sqrt{s_0}$ is the threshold energy and equals the sum of
the $H_c$ mass, the $\bar{D}$ mass, and 2 times the up-quark mass.
The values of the parameters, $a_1$, $b_1$, $c_1$, $a_2$, $b_2$,
and $c_2$, are listed in Tables 1-3. The parametrization as a
function of $\sqrt{s}$ at a given temperature has a peak. In the
three tables we also give the quantities $d_0$ and $\sqrt{s_{\rm
z}}$, where $d_0$ is the separation between the peak's location on
the $\sqrt{s}$-axis and the threshold energy, and $\sqrt{s_{\rm
z}}$ is the square root of the Mandelstam variable at which the
cross section is 1/100 of the peak cross section. We note that Eq.
(20) is valid for $\sqrt{s_0} \leq \sqrt {s} \leq 11$ GeV.

Having obtained the unpolarized cross sections for the 15 reaction
channels, we need to evaluate their influence on the charmonium
nuclear modification factor. We thus calculate the dissociation
rate of charmonium in the interaction with a pion in hadronic
matter, 
\begin{eqnarray}
n_{\pi}\langle v_{\rm rel} \sigma^{\rm unpol} \rangle=
\frac{3}{4\pi^2}
\int^\infty_0\int^1_{-1}d|\vec{k}|d\cos\theta\vec{k}^2v_{\rm rel}
\sigma^{\rm unpol}f(\vec{k}),
\end{eqnarray}
where $n_{\pi}$ is the $\pi$ number density, $v_{\rm rel}$ is the 
relative velocity of the pion and the charmonium, $\theta$ is
the angle between the $\pi$ momentum $\vec{k}$ and the 
charmonium momentum, $f(\vec{k})$ is the Bose-Einstein\
distribution that pions obey, and the thermal average
$<v_{\rm{rel}}\sigma^{\rm{unpol}}>$ is obtained from $f(\vec{k})$
\cite{JSX}. The dissociation rates are
plotted in Figs. 5-7 as functions of charmonium momentum. The rate
shown in Fig. 5 is obtained from the sum of the cross sections for
$\pi+J/\psi\to D^++X$, $\pi+J/\psi\to D^0+X$, $\pi+J/\psi\to
D_s^++X$, $\pi+J/\psi\to D^{*+}+X$, and $\pi+J/\psi\to D^{*0}+X$.
The rate shown in Fig. 6 (7) is obtained from the cross section
for $\pi\psi'\to D^+X+D^0X+D_s^+X+D^{*+}X+D^{*0}X$ ($\pi\chi_c\to
D^+X+D^0X+D_s^+X+D^{*+}X+D^{*0}X$). Given a charmonium momentum 
the rate increases with increasing temperature. Given a
temperature the rate increases with increasing charmonium
momentum. At $T/T_{\rm{c}}=$ 0.95 the rate is about 0.0108 $c$/fm,
0.0142 $c$/fm, and 0.0149 $c$/fm for the $\pi+J/\psi$, $\pi+\psi'$, and
$\pi+\chi_c$ reactions, respectively, when the charmonium momentum
is 20 GeV/$c$. Because the nuclear modification factor is the
inverse of the exponential function of the dissociation rate
\cite{Xu2002}, the reactions give a decreasing charmonium nuclear
modification factor with increasing transverse momentum. This is in
sharp contrast 
with the increasing nuclear modification factor due to 
nuclear shadowing. Decreasing nuclear modification factors are also
caused by other factors, for example, decreasing dissociation
temperature with increasing $J/\psi$ velocity \cite{LRW} and
radiative energy loss of a gluon before gluon
fragmentation into a $J/\psi$ meson.

The relative velocity depends on the pion and charmonium masses.
Since the $\psi'$ mass is very close to the $\chi_c$ mass for
$0.6 \leq T/T_{\rm c} < 1$ \cite{ZX}, 
the relative velocity of $\pi$ and
$\psi'$ is almost the same as the one of $\pi$ and $\chi_c$.
The cross sections for
$\pi+\psi' \to D^++X$, $\pi+\psi' \to D^0+X$, $\pi+\psi' \to
D_s^++X$, $\pi+\psi' \to D^{*+}+X$, and $\pi+\psi' \to D^{*0}+X$
are close to those for 
$\pi+\chi_c\to D^++X$, $\pi+\chi_c\to D^0+X$, $\pi+\chi_c\to
D_s^++X$, $\pi+\chi_c\to D^{*+}+X$, and $\pi+\chi_c\to D^{*0}+X$,
respectively.
To account for this, in Table 4 we show the ratio of the
difference of the cross section for 
$\pi+\psi' \to D^++X$ ($\pi+\psi' \to D^0+X$, $\pi+\psi' \to
D_s^++X$, $\pi+\psi' \to D^{*+}+X$, $\pi+\psi' \to D^{*0}+X$)
and the one for
$\pi+\chi_c\to D^++X$ ($\pi+\chi_c\to D^0+X$, $\pi+\chi_c\to
D_s^++X$, $\pi+\chi_c\to D^{*+}+X$, $\pi+\chi_c\to D^{*0}+X$)
at $\sqrt{s}=9$ GeV to the former. The absolute values of the
ratios are less than 0.25 or even close to 0.
According to Eq. (21), the
dissociation rate of $\psi'$ in the interaction with the pion is
similar to the dissociation rate of $\chi_c$.

The meson masses and quark-antiquark relative-motion wave
functions determined by the Schr\"odinger equation with the
potential in Eq. (14) depend on temperature. Consequently,
$p^\prime_{c\rm min}$ given in Eq. (6), 
$\mid \vec{p}_q^{~\prime} \mid_0$ determined by Eq. (3), and the
transition amplitude ${\cal M}_{\rm fi}$ depend on temperature.
The cross section given by Eq. (7) may not result in a 
monotonously increasing or decreasing function of temperature.
However, the dissociation rate at a given $\sqrt s$ is a 
monotonously increasing function of temperature. The reason is
as follows. With increasing temperature the meson masses
decrease \cite{ZX},
the pion distribution function thus increases, and the relative
velocity of the pion and the charmonium increases as well.
Because the dissociation rate is proportional to the relative
velocity and the pion distribution function, the rate increases
with increasing temperature.

The potential in Eq. (14) is valid at $T<T_{\rm c}$. Solving
the Schr\"odinger equation with the potential, we find that
$J/\psi$, $\psi'$, $\chi_c (1P)$, and $\chi_c (2P)$ are not
dissolved in hadronic matter. Similar to Refs. \cite{BT} and
\cite{GI}, the experimentally observed states, $\psi(4160)$ and 
$\psi(4415)$, can be individually interpreted as the 3$^3S_1$
and 4$^3S_1$ states that satisfy the Schr\"odinger equation.
When the temperature is larger than 0.97$T_{\rm c}$ and
0.87$T_{\rm c}$, respectively, $\psi(4160)$ and $\psi(4415)$
are dissolved.

\vspace{0.5cm}
\leftline{\bf 4. Summary}
\vspace{0.5cm}

We have presented a mechanism for large-momentum
charmonium dissociation. In this mechanism the collision between a
light meson and a charmonium produces two quarks and two
antiquarks; the charm quark fragments into a charmed meson, and
the other three constituents give rise to two or more mesons. The
cross section for the reaction obeying the mechanism is derived
from the transition amplitude and is calculated from the
temperature-dependent quark potential. The temperature dependence
of the potential, the meson masses, and the mesonic
quark-antiquark relative-motion wave functions lead to the
temperature dependence of the cross section. From the cross
section the dissociation rate of charmonium in the interaction
with pions increases with increasing temperature and/or increasing
charmonium momentum, and reaches 0.0108 $c$/fm, 0.0142 $c$/fm, and
0.0149 $c$/fm for the $\pi + J/\psi$, $\pi + \psi^\prime$, and
$\pi + \chi_c$ reactions, respectively, at $T/T_{\rm c}=0.95$ and
at the momentum 20 GeV/$c$. These rates are small, but already 
nonnegligible. In future work we will include collisions of charmonia 
with $K, \eta, \rho, K^*$ and $\phi$ mesons. The
mechanism for large-momentum charmonium dissociation is in contrast
to the quark-interchange mechanism for low-momentum charmonium
dissociation. The decreasing nuclear modification factor of the
prompt $J/\psi$ with increasing momentum is caused not only by
$\pi J/\psi \to D^+X+D^0X+D_s^+X+D^{*+}X+D^{*0}X$,
but also by other factors, for example, decreasing dissociation
temperature with increasing $J/\psi$ velocity \cite{LRW}
and radiative energy loss of a gluon before gluon fragmentation into
a $J/\psi$ meson.

\vspace{0.5cm}
\leftline{\bf Acknowledgements}
\vspace{0.5cm}
We thank anonymous referees whose comments led us to improve  
our work substantially.  
This work was supported by the National Natural Science Foundation of China 
under Grant No. 11175111.

\newpage
\begin{figure}[htbp]
  \centering
    \includegraphics[scale=0.6]{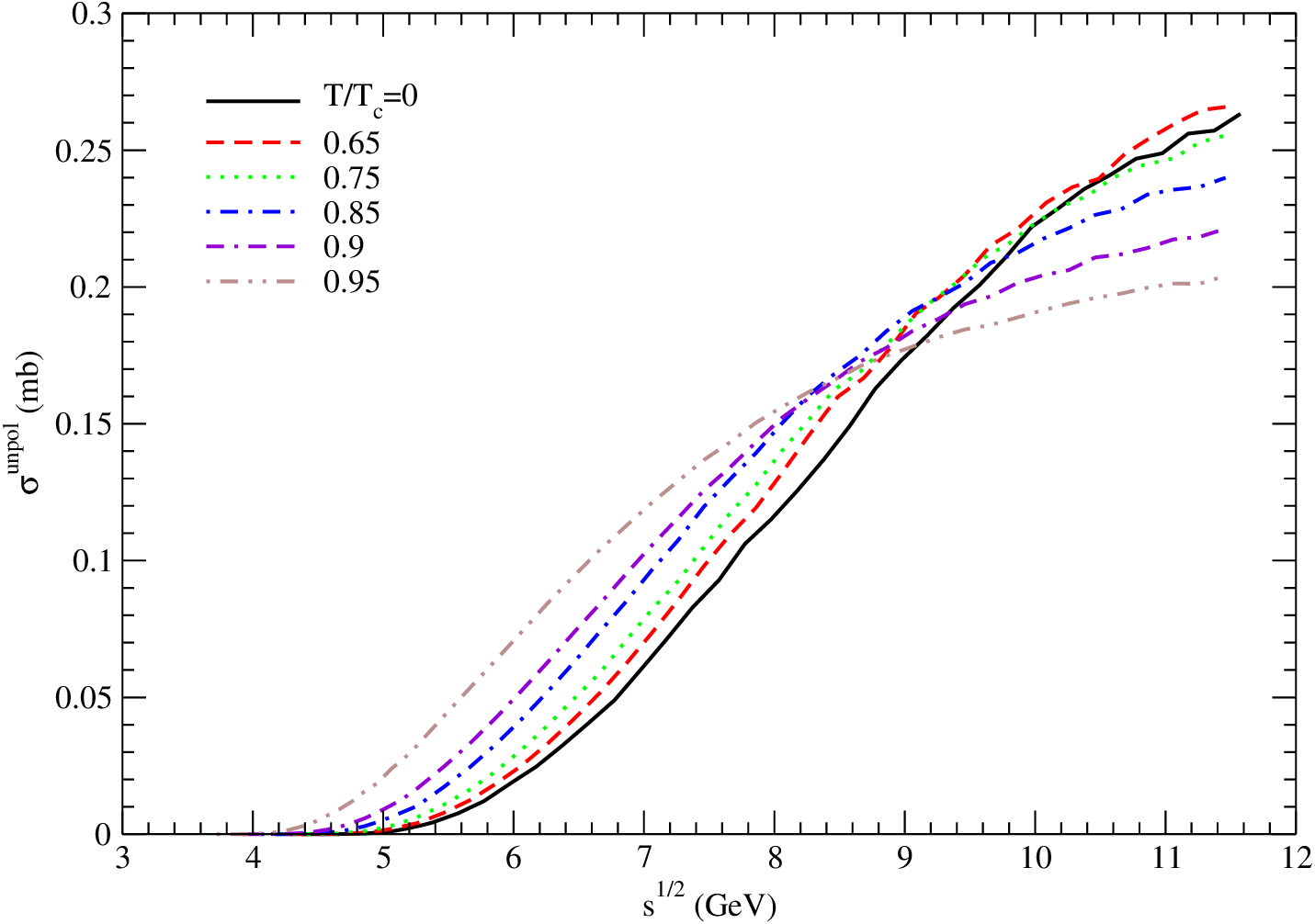}%
\caption{Cross sections for $\pi + J/\psi \to D^+ + X$ at various
temperatures.} \label{fig1}
\end{figure}

\newpage
\begin{figure}[htbp]
  \centering
    \includegraphics[scale=0.6]{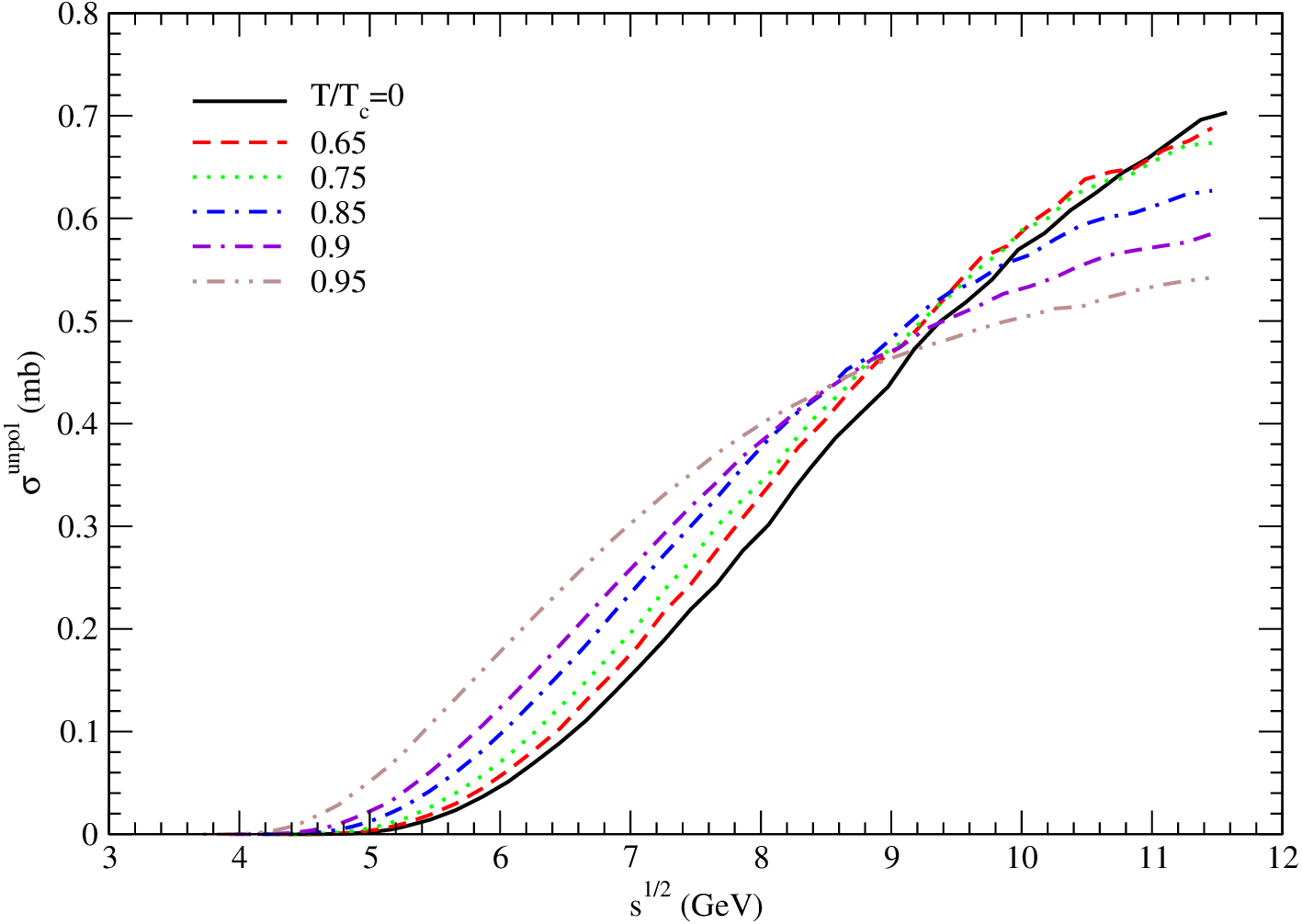}%
\caption{Cross sections for $\pi + J/\psi \to D^0 + X$ at various
temperatures.} \label{fig2}
\end{figure}

\newpage
\begin{figure}[htbp]
  \centering
    \includegraphics[scale=0.6]{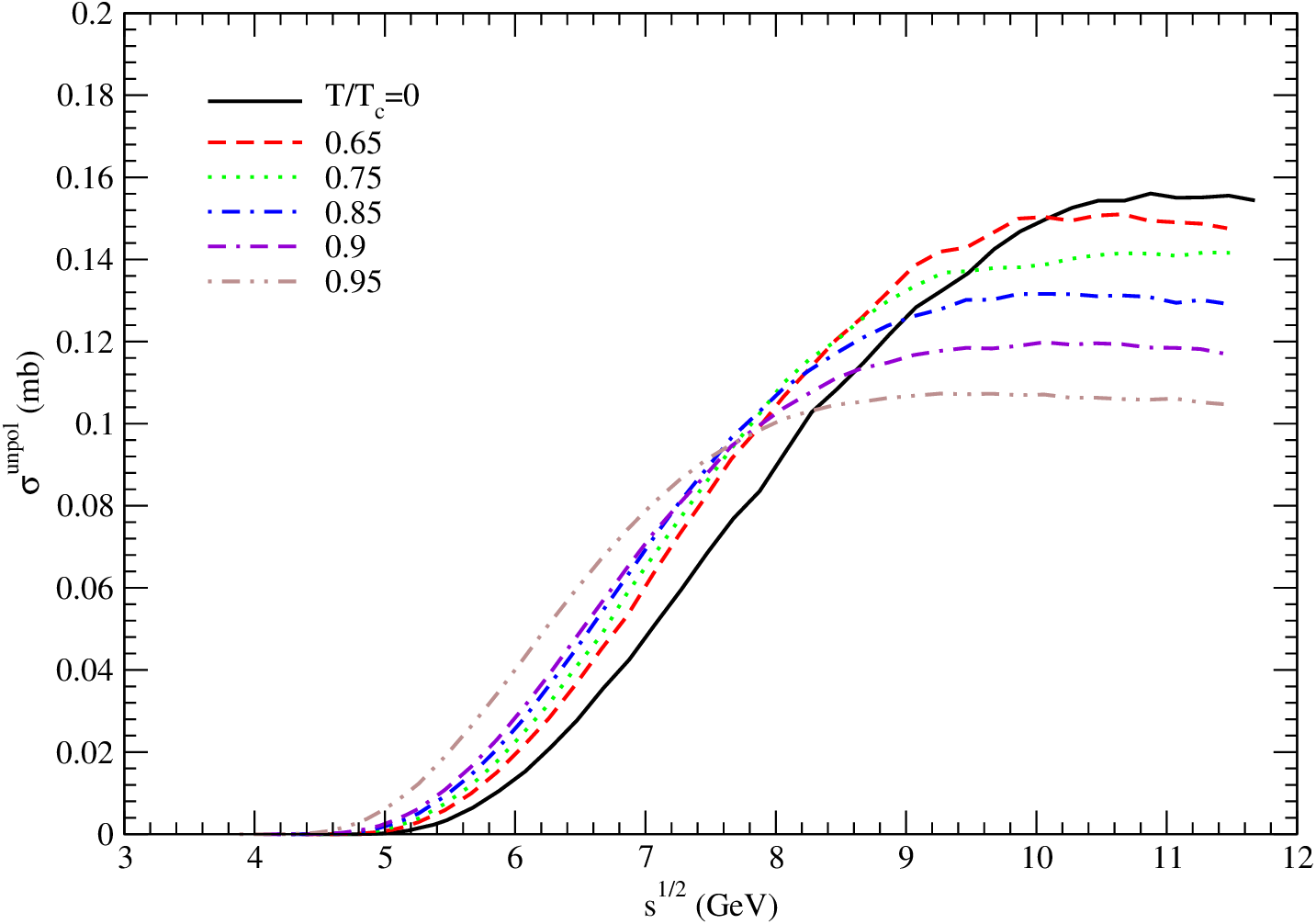}%
\caption{Cross sections for $\pi + J/\psi \to D^+_s + X$ at
various temperatures.} \label{fig3}
\end{figure}

\newpage
\begin{figure}[htbp]
  \centering
    \includegraphics[scale=0.6]{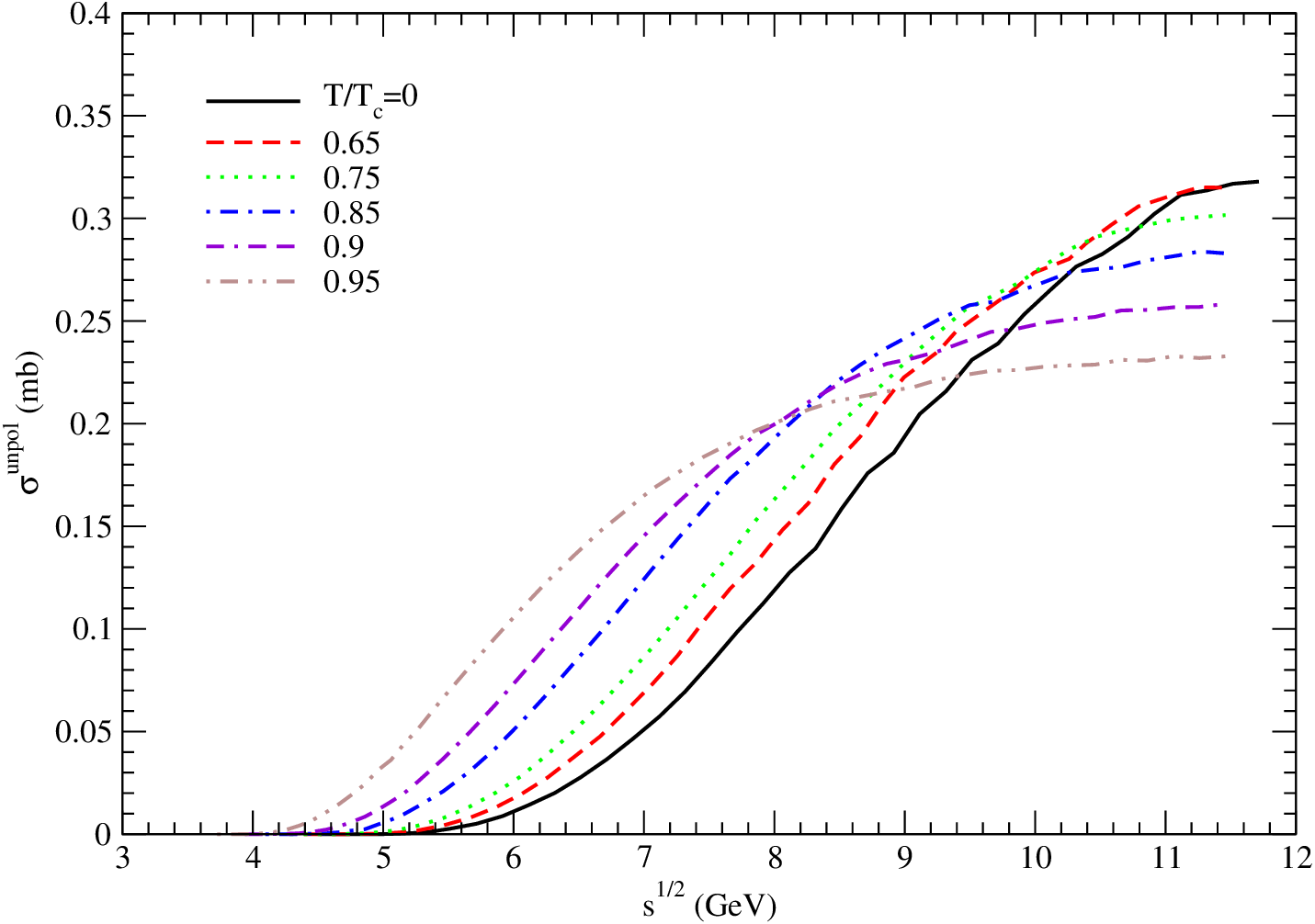}%
\caption{Cross sections for $\pi + J/\psi \to D^{*+} + X$ at
various temperatures.} \label{fig4}
\end{figure}

\newpage
\begin{figure}[htbp]
  \centering
    \includegraphics[scale=0.6]{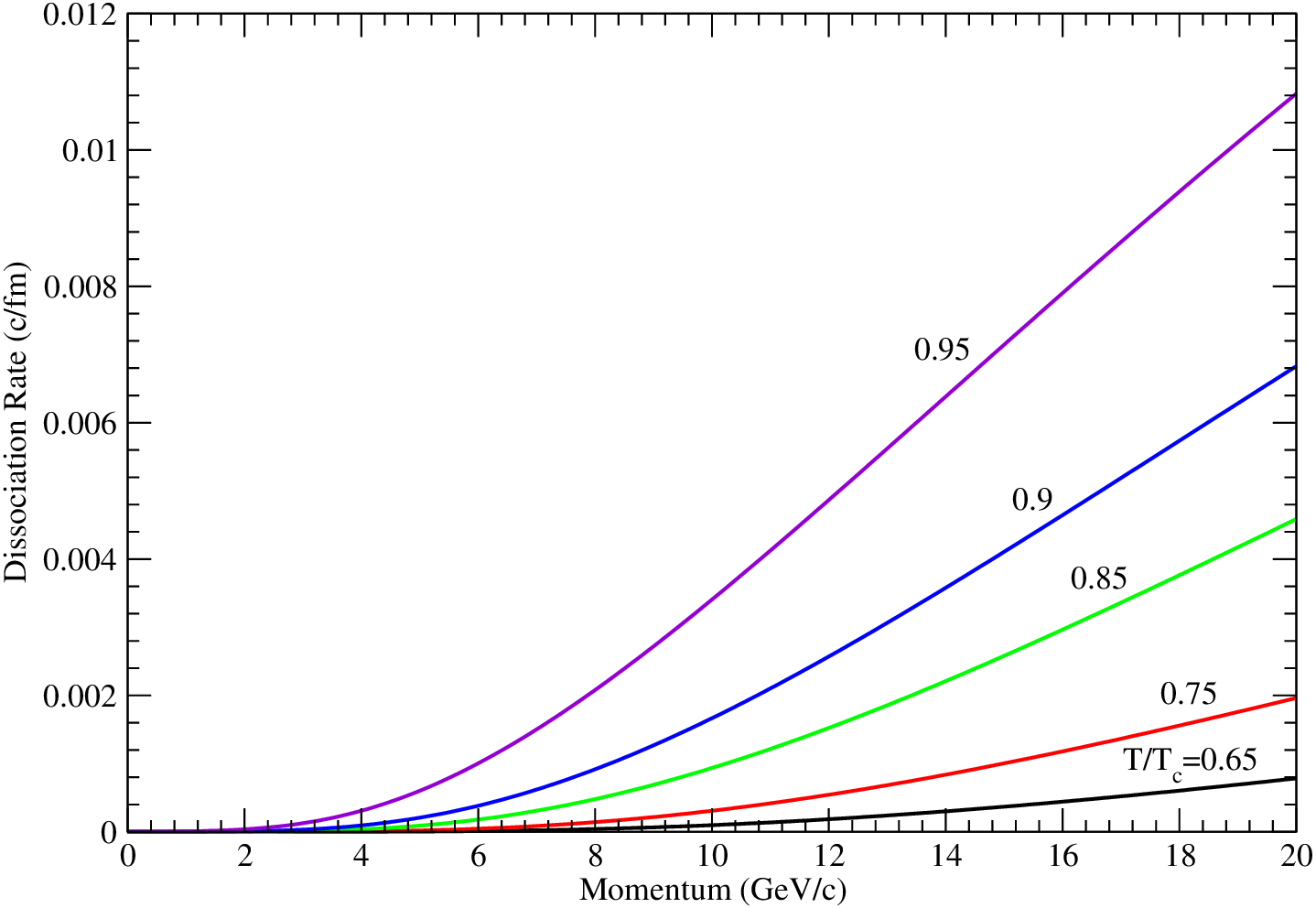}%
\caption{Dissociation rate of $J/\psi$ with $\pi$ versus $J/\psi$
momentum at various temperatures.} \label{fig5}
\end{figure}

\newpage
\begin{figure}[htbp]
  \centering
    \includegraphics[scale=0.6]{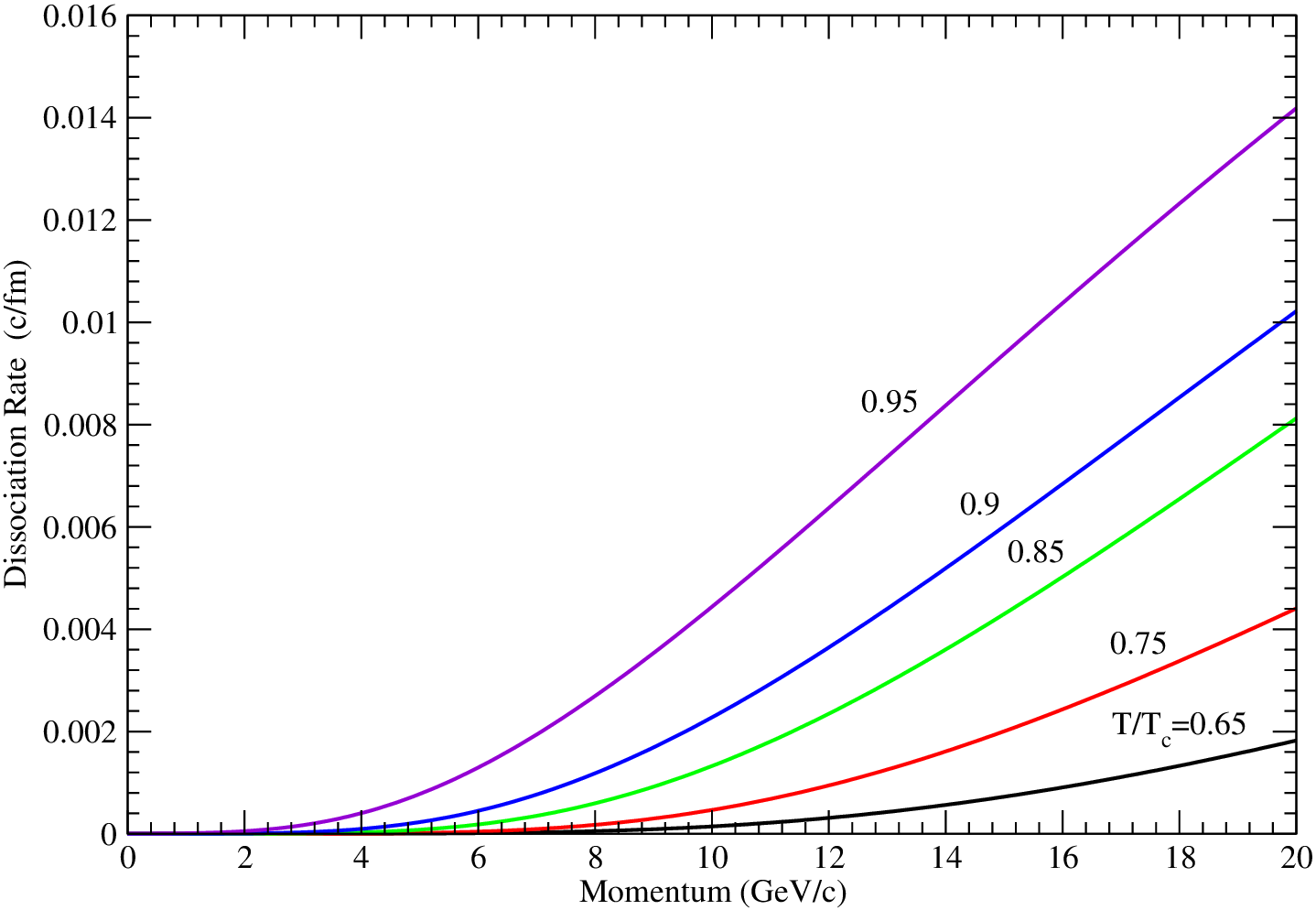}%
\caption{Dissociation rate of $\psi^\prime$ with $\pi$ versus
$\psi^\prime$ momentum at various temperatures.} \label{fig6}
\end{figure}

\newpage
\begin{figure}[htbp]
  \centering
    \includegraphics[scale=0.6]{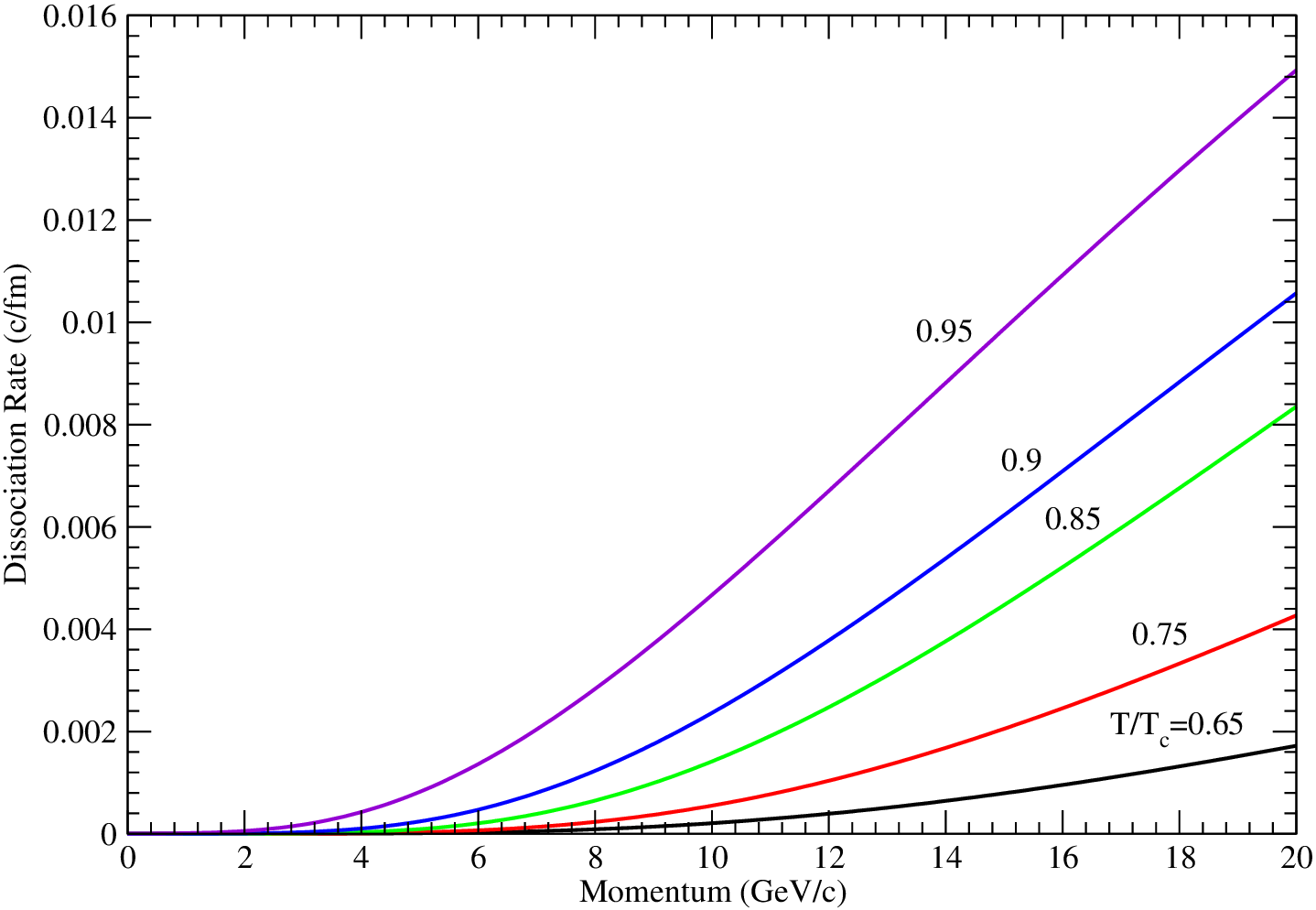}%
\caption{Dissociation rate of $\chi_c$ with $\pi$ versus $\chi_c$
momentum at various temperatures.} \label{fig7}
\end{figure}

\newpage
\begin{figure}[htbp]
  \centering
    \includegraphics[scale=0.6]{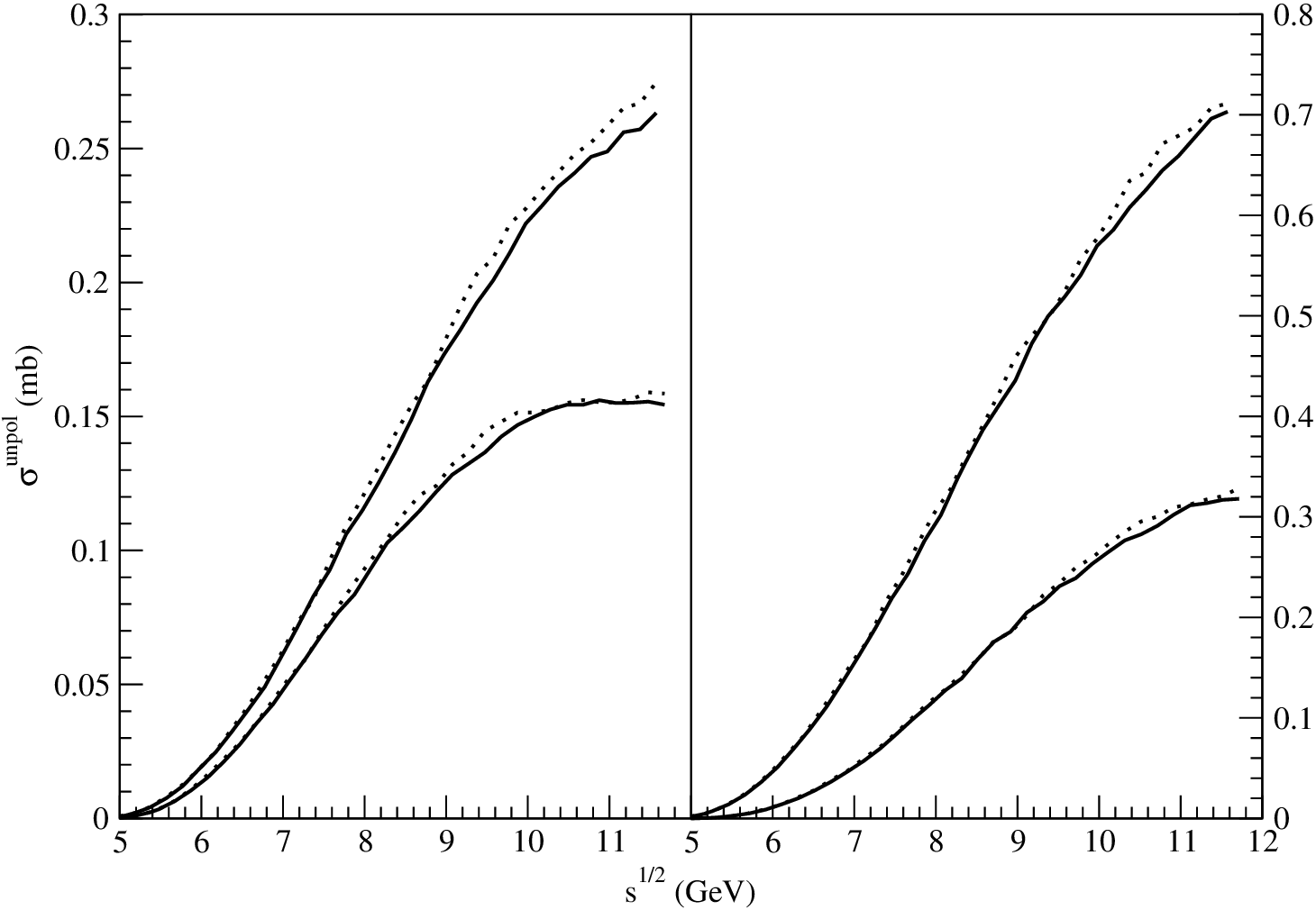}%
\caption{Solid and dotted curves stand for the unpolarized cross
sections corresponding to Eqs. (14) and (19), respectively.
Left (right) panel: the upper and lower pairs of curves correspond to
$\pi+J/\psi\to D^++X$ and $\pi+J/\psi\to D_s^++X$
($\pi+J/\psi\to D^0+X$ and $\pi+J/\psi\to D^{*+}+X$), respectively.}
\label{fig8}
\end{figure}

\begin{table}
\centering \caption{Quantities relevant to the cross sections for
the $\pi J/\psi$ dissociation reactions. $a_1$ and $a_2$ are in
units of mb; $b_1$, $b_2$, $d_0$, and $\sqrt{s_{\rm z}}$ are in
units of GeV; $c_1$ and $c_2$ are dimensionless.} \tabcolsep=5pt
\begin{tabular}{llllllllll}
  \hline
  Reactions & $T/T_{\rm c}$ & $a_1$ & $b_1$ & $c_1$ & $a_2$ & $b_2$ & $c_2$ &$d_0$ & $\sqrt{s_{\rm z}}$\\
  \hline
  $\pi+J/\psi\to D^++X$
  & 0     & 0.02    & 2.8   & 4.9   & 0.26   & 7.4   & 4.7 & 7.36 & 27.43\\
  & 0.65  & 0.01    & 2.7   & 4.8   & 0.26   & 7.4   & 4   & 7.38 & 29.19\\
  & 0.75  & 0.01    & 2.5   & 5.9   & 0.25   & 7.3   & 3.8 & 7.29 & 29.39\\
  & 0.85  & 0.021   & 2.9   & 4.6   & 0.24   & 7.4   & 3.9 & 7.33 & 29.26\\
  & 0.9   & 0.07    & 3.7   & 4.1   & 0.21   & 8.5   & 4.6 & 7.87 & 30.53\\
  & 0.95  & 0.04    & 2.7   & 4.5   & 0.2    & 7.2   & 3.8 & 7.05 & 28.54\\
  \hline
  $\pi+J/\psi\to D^0+X$
  & 0     & 0.015   & 2.16  & 6.3   & 0.69   & 7.68  & 3.9 & 7.68 & 30.54\\
  & 0.65  & 0.03    & 2.7   & 5.2   & 0.68   & 7.6   & 4   & 7.59 & 29.86\\
  & 0.75  & 0.06    & 3.2   & 4.6   & 0.67   & 7.9   & 4   & 7.81 & 30.77\\
  & 0.85  & 0.028   & 2.5   & 6.4   & 0.61   & 7.2   & 3.6 & 7.19 & 29.59\\
  & 0.9   & 0.064   & 2.93  & 4.5   & 0.57   & 7.38  & 3.8 & 7.27 & 29.38\\
  & 0.95  & 0.08    & 2.5   & 4.7   & 0.53   & 6.9   & 3.9 & 6.83 & 27.22\\
  \hline
  $\pi+J/\psi\to D_s^++X$
  & 0     & 0.02    & 3.5   & 4.7   & 0.15   & 6.9   & 4.1 & 6.61 & 27.34\\
  & 0.65  & 0.02    & 2.7   & 5.2   & 0.15   & 6.1   & 4.9 & 5.98 & 22.99\\
  & 0.75  & 0.05    & 4     & 4.5   & 0.12   & 7.5   & 3.9 & 6.22 & 29.36\\
  & 0.85  & 0.018   & 3.1   & 6.5   & 0.132  & 6.2   & 3.9 & 5.98 & 25.25\\
  & 0.9   & 0.04    & 3.5   & 5.3   & 0.11   & 6.8   & 4.5 & 5.93 & 25.47\\
  & 0.95  & 0.05    & 3.3   & 5.3   & 0.1    & 7.1   & 4.9 & 6.11 & 25.41\\
  \hline
  $\pi+J/\psi\to D^{*+}+X$
  & 0     & 0.027   & 3.5   & 4.9   & 0.31   & 7.5   & 5.3 & 7.39 & 26.63\\
  & 0.65  & 0.04    & 3.3   & 5.4   & 0.3    & 7     & 6   & 6.86 & 23.93\\
  & 0.75  & 0.01    & 2.6   & 8     & 0.31   & 7.2   & 4   & 7.2  & 28.51\\
  & 0.85  & 0.01    & 1.7   & 14    & 0.29   & 6.3   & 3.9 & 6.3  & 25.57\\
  & 0.9   & 0.06    & 3     & 4.5   & 0.25   & 7     & 4.2 & 6.67 & 26.88\\
  & 0.95  & 0.06    & 2.58  & 4.5   & 0.232  & 6.7   & 3.8 & 6.48 & 26.81\\
  \hline
\end{tabular}
\end{table}
\begin{table}
\centering \caption{The same as Table 1 except for $\pi\psi'$.}
\tabcolsep=5pt
\begin{tabular}{llllllllll}
  \hline
  Reactions & $T/T_{\rm c}$ & $a_1$ & $b_1$ & $c_1$ & $a_2$ & $b_2$ & $c_2$ &$d_0$ & $\sqrt{s_{\rm z}}$\\
  \hline
  $\pi+\psi'\to D^++X$
  & 0     & 0.013   & 1.8   & 8.9   & 0.44   & 6.6   & 3.9 & 6.6  & 26.86\\
  & 0.65  & 0.209   & 4     & 13.1  & 0.76   & 9.2   & 4.7 & 9.18 & 32.97\\
  & 0.75  & 0.13    & 3.4   & 16    & 0.6    & 7.3   & 5.5 & 7.29 & 25.4 \\
  & 0.85  & 0.11    & 3.4   & 9     & 0.4    & 7.1   & 5.5 & 6.92 & 24.65\\
  & 0.9   & 0.067   & 3.2   & 6.7   & 0.3    & 7     & 4.5 & 6.78 & 26.16\\
  & 0.95  & 0.04    & 2.9   & 5     & 0.24   & 7     & 3.1 & 6.78 & 30.48\\
  \hline
  $\pi+\psi'\to D^0+X$
  & 0     & 0.03    & 1.6   & 15    & 1.16   & 6.8   & 3.8 & 6.8  & 27.84\\
  & 0.65  & 0.001   & 0.7   & 5     & 1.8    & 6     & 6.8 & 6    & 20.21\\
  & 0.75  & 0.7     & 4.1   & 11    & 3.22   & 14.4  & 4.1 & 14.4 & 52.08\\
  & 0.85  & 0.12    & 2.9   & 12.7  & 1.06   & 6.3   & 5.5 & 6.29 & 22.35\\
  & 0.9   & 0.13    & 3     & 7.5   & 0.81   & 6.9   & 4.4 & 6.82 & 26.1 \\
  & 0.95  & 0.13    & 3     & 4     & 0.61   & 6.9   & 3.9 & 6.54 & 27.14\\
  \hline
  $\pi+\psi'\to D_s^++X$
  & 0     & 0.012   & 2.2   & 8.4   & 0.27   & 5.9   & 3.8 & 5.9  & 24.84\\
  & 0.65  & 0.27    & 3.9   & 17    & 0.42   & 9     & 4.5 & 8.99 & 32.99\\
  & 0.75  & 0.15    & 3.3   & 27    & 0.34   & 6     & 6.7 & 5.98 & 20.3 \\
  & 0.85  & 0.1     & 3.2   & 20    & 0.249  & 6     & 6.7 & 5.95 & 20.18\\
  & 0.9   & 0.071   & 3.2   & 14.5  & 0.175  & 6     & 6.2 & 5.8  & 20.6 \\
  & 0.95  & 0.06    & 3.1   & 8.7   & 0.13   & 6.3   & 4.9 & 5.7  & 23.08\\
  \hline
  $\pi+\psi'\to D^{*+}+X$
  & 0     & 0.009   & 1.7   & 12.1  & 0.52   & 6.8   & 4.2 & 6.8  & 26.87\\
  & 0.65  & 0.4     & 4.3   & 18    & 1.98   & 14.4  & 4.4 & 14.4 & 50.68\\
  & 0.75  & 0.23    & 3.8   & 19    & 0.73   & 7.8   & 6   & 7.78 & 26.09\\
  & 0.85  & 0.116   & 3.1   & 15    & 0.5    & 6.1   & 5.9 & 6.05 & 21.28\\
  & 0.9   & 0.1     & 2.8   & 7.4   & 0.37   & 6     & 5.9 & 5.83 & 20.81\\
  & 0.95  & 0.12    & 3.2   & 4     & 0.25   & 7     & 3.8 & 5.67 & 27.55\\
  \hline
\end{tabular}
\end{table}
\begin{table}
\centering \caption{The same as Table 1 except for $\pi\chi_c$.}
\tabcolsep=5pt
\begin{tabular}{llllllllll}
  \hline
  Reactions & $T/T_{\rm c}$ & $a_1$ & $b_1$ & $c_1$ & $a_2$ & $b_2$ & $c_2$ &$d_0$ & $\sqrt{s_{\rm z}}$\\
  \hline
  $\pi+\chi_c\to D^++X$
  & 0     & 0.07    & 3     & 4.3   & 0.41  & 7.6    & 4.8 & 7.45 & 27.8 \\
  & 0.65  & 0.03    & 3.4   & 14    & 0.53  & 7.2    & 4.2 & 7.19 & 27.95\\
  & 0.75  & 0.11    & 3.8   & 9.6   & 0.5   & 7.7    & 4.4 & 7.48 & 28.92\\
  & 0.85  & 0.12    & 3.4   & 7.8   & 0.39  & 6.8    & 6.6 & 6.51 & 22.27\\
  & 0.9   & 0.1     & 3.5   & 5.7   & 0.29  & 7.3    & 5.2 & 6.75 & 25.54\\
  & 0.95  & 0.1     & 3.7   & 4     & 0.25  & 9.6    & 3   & 8.59 & 40.94\\
  \hline
  $\pi+\chi_c\to D^0+X$
  & 0     & 0.06    & 2.3   & 5     & 1.04  & 7.2    & 3.8 & 7.19 & 29.22\\
  & 0.65  & 0.14    & 4     & 9     & 1.39  & 7.9    & 4   & 7.76 & 30.85\\
  & 0.75  & 0.42    & 4.1   & 8.6   & 1.58  & 10.1   & 3.9 & 10.02& 38.6 \\
  & 0.85  & 0.2     & 3.2   & 8.7   & 1.05  & 6.7    & 5.6 & 6.59 & 23.35\\
  & 0.9   & 0.25    & 3.6   & 5.9   & 0.81  & 8      & 4.4 & 7.56 & 29.57\\
  & 0.95  & 0.12    & 3     & 4     & 0.64  & 6.8    & 3.7 & 6.46 & 27.41\\
  \hline
  $\pi+\chi_c\to D_s^++X$
  & 0     & 0.04    & 2.6   & 5.2   & 0.25  & 6.5    & 4.3 & 6.4  & 25.59\\
  & 0.65  & 0.1     & 4.1   & 7     & 0.27  & 6.5    & 4.5 & 5.52 & 24.64\\
  & 0.75  & 0.11    & 3.5   & 15    & 0.3   & 6.3    & 5.1 & 5.89 & 23.16\\
  & 0.85  & 0.13    & 3.4   & 13.3  & 0.23  & 6.3    & 7.5 & 5.88 & 20.2 \\
  & 0.9   & 0.11    & 3.4   & 11.3  & 0.17  & 6.4    & 8.3 & 5.82 & 19.7 \\
  & 0.95  & 0.065   & 3     & 9.9   & 0.14  & 6.1    & 5.6 & 5.78 & 21.42\\
  \hline
  $\pi+\chi_c\to D^{*+}+X$
  & 0     & 0.03    & 2.6   & 6.1   & 0.49  & 7.2    & 4.5 & 7.19 & 27.42\\
  & 0.65  & 0.13    & 4.4   & 10.8  & 0.6   & 7.8    & 4.6 & 7.29 & 28.87\\
  & 0.75  & 0.2     & 4.1   & 12    & 0.62  & 8.1    & 4.8 & 7.85 & 29.26\\
  & 0.85  & 0.21    & 3.5   & 9     & 0.47  & 6.7    & 7.5 & 6.25 & 21.12\\
  & 0.9   & 0.11    & 3.1   & 7     & 0.36  & 6.6    & 4.6 & 6.26 & 24.66\\
  & 0.95  & 0.13    & 3.2   & 4     & 0.27  & 7.5    & 3.7 & 6.22 & 29.69\\
  \hline
  \end{tabular}
\end{table}
\begin{table}
\centering \caption{Ratio of the difference of the cross sections for
$\pi + \psi' \to H_c +X$ and for $\pi + \chi_c \to H_c +X$ to the cross
section for $\pi + \psi' \to H_c +X$ at $\sqrt{s}=9$ GeV.}
\begin{tabular}{|l|l|l|l|l|l|}
  \hline
  ~ & $T/T_{\rm c}=0.65$ & $T/T_{\rm c}=0.75$ & $T/T_{\rm c}=0.85$ 
& $T/T_{\rm c}=0.9$ & $T/T_{\rm c}=0.95$\\
  \hline
  $D^++X$
  & 0.219  & 0.102  & -0.0048  & -0.0144  & -0.047 \\
  \hline
  $D^0+X$
  & 0.212  & 0.103  & 0.002  & -0.014  & -0.042 \\
  \hline
  $D_s^++X$
  & 0.241  & 0.107  & -0.0063  & -0.0183  & -0.045 \\
  \hline
  $D^{*+}+X$
  & 0.217  & 0.108  & -0.0103  & -0.0136  & -0.045 \\
  \hline
  $D^{*0}+X$
  & 0.217  & 0.108  & -0.0103  & -0.0136  & -0.045 \\
  \hline
  \end{tabular}
\end{table}

\begin{thebibliography}{00}
\bibitem{Peskin}M. E. Peskin, Nucl. Phys. B {\bf 156}, 365 (1979); G. Bhanot 
and M. E. Peskin, Nucl. Phys. B {\bf 156}, 391 (1979).
\bibitem{KS}D. Kharzeev and H. Satz, Phys. Lett. B {\bf 334}, 155 (1994).
\bibitem{AGGA}F. Arleo, P. B. Gossiaux, T. Gousset, and J. Aichelin, Phys. Rev.
D {\bf 65}, 014005 (2001).
\bibitem{MM}S. G. Matinyan and B. M\"{u}ller, Phys. Rev. C {\bf 58}, 2994
(1998). 
\bibitem{HG}K. L. Haglin and C. Gale, Phys. Rev. C {\bf 63}, 065201 (2001).
\bibitem{LK}Z. Lin and C. M. Ko, Phys. Rev. C {\bf 62}, 034903 (2000).
\bibitem{OSL}Y. S. Oh, T. S. Song, and S. H. Lee, Phys. Rev. C {\bf 63}, 034901
(2001).
\bibitem{NNR}F. S. Navarra, M. Nielsen, and M. R. Robilotta, Phys. Rev. C {\bf 
64}, 021901(R) (2001).
\bibitem{MPPR}L. Maiani, F. Piccinini, A. D. Polosa, and V. Riquer, Nucl. Phys.
A {\bf 741}, 273 (2004).
\bibitem{BG}A. Bourque and C. Gale, Phys. Rev. C {\bf 80}, 015204 (2009).
\bibitem{MBQ}K. Martins, D. Blaschke, and E. Quack, Phys. Rev. C 51, 2723 
(1995).
\bibitem{WSB}C.-Y. Wong, E. S. Swanson, and T. Barnes, Phys. Rev. C 65, 014903
(2001).
\bibitem{BSWX}T. Barnes, E. S. Swanson, C.-Y. Wong, and X.-M. Xu, Phys. Rev. C
68, 014903 (2003).
\bibitem{ZX}J. Zhou and X.-M. Xu, Phys. Rev. C 85, 064904 (2012).
\bibitem{JSX}S.-T. Ji, Z.-Y. Shen, and X.-M. Xu, J. Phys. G 42, 095110 (2015).
\bibitem{PHENIX}A. Adare et al. (PHENIX Collaboration), Phys. Rev. Lett. 98, 
232301 (2007);
Phys. Rev. Lett. 101, 122301 (2008);
Phys. Rev. C 84, 054912 (2011).
\bibitem{STAR}B. I. Abelev et al. (STAR Collaboration), Phys. Rev. C 80, 
041902 (2009);
L. Adamczyk et al. (STAR Collaboration), Phys. Lett. B 722, 55 (2013);
Phys. Rev. C 90, 024906 (2014).
\bibitem{CMS}S. Chatrchyan et al. (CMS Collaboration), JHEP 05, 063 (2012);
P. Shukla, for the CMS Collaboration, arXiv:1405.3810, talk given
at International Conference on Matter at Extreme Conditions, 15-17
January 2014, Kolkata, India.
\bibitem{ATLAS}S. T. Araya, for the ATLAS Collaboration, talk given 
at the 8th International Conference on Hard and Electromagnetic Probes of
High-Energy Nuclear Collisions, 23-27 September 2016, Wuhan, China.
\bibitem{LJX}F.-R. Liu, S.-T. Ji, and X.-M. Xu, J. Korean Phys.
Soc. 69, 472 (2016).
\bibitem{Xu1999}X.-M. Xu, Nucl. Phys. A 658, 165 (1999).
\bibitem{DR}X. Du and R. Rapp, Nucl. Phys. A 943, 147 (2015).
\bibitem{Arneodo}M. Arneodo, Phys. Rep. 240, 301 (1994).
\bibitem{EKS}K. J. Eskola, V. J. Kolhinen, and C. A. Salgado, Eur. Phys. J. C 
9, 61 (1999); K. J. Eskola, V. J. Kolhinen, and P. V. Ruuskanen, 
hep-ph/9802350.
\bibitem{LW}S.-Y. Li and X.-N. Wang, Phys. Lett. B 527, 85 (2002).
\bibitem{XKSW}X.-M. Xu, D. Kharzeev, H. Satz, and X.-N. Wang, Phys. Rev. C 53, 
3051 (1996).
\bibitem{ZXXZ}K. Zhou, N. Xu, Z. Xu, and P. Zhuang, Phys. Rev. C 89, 054911 
(2014).
\bibitem{LRW}H. Liu, K. Rajagopal, and U. A. Wiedemann, Phys. Rev. Lett. 98,
182301 (2007).
\bibitem{KKKS}T. Kneesch, B. A. Kniehl, G. Kramer,
and I. Schienbein, Nucl. Phys. B 799, 34 (2008).
\bibitem{KK}B. A. Kniehl and G. Kramer, Phys. Rev. D 74, 037502 (2006).
\bibitem{BT}W. Buchm\"{u}ller and S.-H. H. Tye, Phys. Rev. D 24, 132 (1981).
\bibitem{KLP}F. Karsch, E. Laermann, and A. Peikert, Nucl. Phys. B 605, 579 
(2001).
\bibitem{Wong2005}C.-Y. Wong, Phys. Rev. C 72, 034906 (2005).
\bibitem{LPTR}M. Laine, O. Philipsen, M. Tassler, and P. Romatschke, JHEP
03, 054 (2007).
\bibitem{BGVP}N. Brambilla, J. Ghiglieri, A. Vairo, and P. Petreczky, Phys.
Rev. D 78, 014017 (2008).
\bibitem{BKR14}Y. Burnier, O. Kaczmarek, and A. Rothkopf, arXiv:1410.7311.
\bibitem{BKR16}Y. Burnier, O. Kaczmarek, and A. Rothkopf, JHEP 12, 101 (2015).
\bibitem{BBP}A. Bazavov, Y. Burnier, and P. Petreczky, Nucl. Phys. A 932, 117
(2014).
\bibitem{LMSK}S. H. Lee, K. Morita, T. Song, and C. M. Ko, Phys. Rev.
D 89, 094015 (2014).
\bibitem{Satz}H. Satz, Eur. Phys. J. C 75, 193 (2015).
\bibitem{RHS}A. Rothkopf, T. Hatsuda, and S. Sasaki, Phys. Rev. Lett. 108,
162001 (2012); Y. Burnier and A. Rothkopf, Phys. Rev. Lett. 111,
182003 (2013).
\bibitem{BR}Y. Burnier and A. Rothkopf, arXiv:1607.04049.
\bibitem{SX}Z.-Y. Shen and X.-M. Xu, Chin. Phys. C 39, 074103 (2015).
\bibitem{BS}T. Barnes and E. S. Swanson, Phys. Rev. D 46, 131 (1992); E. S. 
Swanson, Ann. Phys. (N.Y.) 220, 73 (1992).
\bibitem{GI}S. Godfrey and N. Isgur, Phys. Rev. D 32, 189 (1985).
\bibitem{Xu2002}X.-M. Xu, Nucl. Phys. A 697, 825 (2002).
\bibitem{PDG}J. Beringer et al., Phys. Rev. D 86, 010001 (2012).
\bibitem{Colton} E. Colton et al., Phys. Rev. D 3, 2028 (1971);
N. B. Durusoy et al., Phys. Lett. B 45, 517 (1973);
W. Hoogland et al., Nucl. Phys. B 126, 109 (1977);
M. J. Losty et al., Nucl. Phys. B 69, 185 (1974).
\bibitem{FF}R. D. Field and R. P. Feynman, Nucl. Phys. B136, 1 (1978).
\end{thebibliography}
\end{document}